\documentclass[reprint,twocolumn,aps,prb,superscriptaddress]{revtex4-2}

\usepackage[T1]{fontenc}
\usepackage[utf8]{inputenc}
\usepackage{lmodern}
\usepackage{graphicx}
\usepackage{amsmath}
\usepackage{xcolor}
\usepackage{float}
\usepackage{hyperref}
\usepackage{braket}
\usepackage{amssymb,amsmath,amsthm, array}
\usepackage{mdframed}
\usepackage{systeme,mathtools}
\usepackage{yfonts}
\usepackage{comment}

\usepackage{amsmath}
   \usepackage{multirow}

\DeclareUnicodeCharacter{0301}{-}

\begin{document}
\renewcommand{\vec}[1]{\mathbf{#1}}
\newcommand{\ii}{\mathrm{i}}
\def\ya#1{{\color{orange}{#1}}}

\title{Two-level atom witness of thermalization of multimode optical fibers}


\author{M. Wanic}
\affiliation{Department of Physics and Medical Engineering, Rzesz\'ow University of Technology, 35-959 Rzesz\'ow, Poland}

\author{R. Khomeriki}
\affiliation{Faculty of Exact and Natural Sciences, Tbilisi State University, Chavchavadze av.3, 0128 Tbilisi, Georgia}

\author{S. Stagraczyński}
\affiliation{Department of Physics and Medical Engineering, Rzesz\'ow University of Technology, 35-959 Rzesz\'ow, Poland}

\author{M. I. Katsnelson}
\affiliation{Radboud University, Institute for Molecules and Materials, Heyendaalseweg 135, 6525AJ Nijmegen, Netherlands}

\author{Z. Toklikishvili}
\affiliation{Faculty of Exact and Natural Sciences, Tbilisi State University, Chavchavadze av.3, 0128 Tbilisi, Georgia}

\author{L. Chotorlishvili}
\affiliation{Department of Physics and Medical Engineering, Rzesz\'ow University of Technology, 35-959 Rzesz\'ow, Poland}

\date{\today}

\begin{abstract}
In the present project, we study the dynamics of the two-level system coupled with the multimode optical system. In particular, we considered a square lattice of optical fibers. We aimed to answer whether we can infer information about the thermalization of optical modes through the thermalization of two-level atoms. After averaging over the set of modes, the dynamic of the two-level system is free of quantum revivals, and that is the signature of thermalization. We showed that the temperature of the two-level system increases with the temperature of optical modes and mean photon number. In the high-temperature limit of optical modes, the temperature of the level system tends to be infinity, and level populations are equal.

\end{abstract}

\maketitle

Since the seminal work by Mark Srednicki \cite{PhysRevE.50.888}, the problem of thermalization of the isolated quantum mechanical systems is the focus of interest. Open quantum systems thermalize due to contact with the thermostat \cite{JPSJ1,JPSJ2}. However, thermalization of the isolated quantum system with many degrees of freedom is a more intriguing process \cite{PhysRevLett.106.010405}. Considerable attention has been paid to the thermalization process in multimode optical fiber systems in the last few years \cite{PhysRevLett.128.213901, PhysRevResearch.3.033219}. In such systems, thermalization is a consequence of nonlinear photon-photon interactions. Moreover, thermalization was observed experimentally in the photonic time-synthetic mesh lattices \cite{marques2023observation}. In the present work, we are interested in the question of whether the quantum two-level system can witness the thermalization process in the multimodal optical fiber system. To answer this question, we consider Jaynes Cummings (JC) atom interacting with the set of optical modes. Since which particular mode atom couples with is inaccessible information, we consider averaging procedure over the set of modes. Typically, the JC system interacting with a coherent quantized field is characterized by quantum revivals in level populations. The quantum revival of level populations is the signature of unitary and time-reversible quantum dynamics. However, in what follows, we will prove the absence of quantum revivals after averaging the atom's dynamic over the ensemble of optical modes. The absence of revivals is the essence of thermalization of the atom. Hence, in our scenario the two-level atom interacting with the ensemble of optical modes thermalizes to a specific temperature $T_a$ and witnesses the thermalization of optical modes. \vspace{0.2cm}\\
Let us present the initial state of the quantum system as the linear superposition of eigenstates $\alpha$ of its Hamiltonian $\hat H$ corresponding to the energies $E_{\alpha}$:
$\ket{\Psi(0)}=\sum\limits_{\alpha}c_{\alpha}\ket{\alpha}$. Then, the expectation value of the operator
$\hat{\mathcal{A}}$ calculated through the evolved in time quantum state  $\ket{\Psi(t)}=e^{-it\hat H}\ket{\Psi(0)}$, reads: $\langle\hat{\mathcal{A}}\rangle=\bra{\Psi(t)}\hat{\mathcal{A}}\ket{\Psi(t)}=\sum\limits_{\alpha,\beta}c^*_\alpha c_\beta(\hat{\mathcal{A}})_{\alpha\beta}e^{-i\omega_{\alpha\beta}t}$. The eigenstate thermalization hypothesis (ETH) argues that \cite{PhysRevA.43.2046} at times $t\gg 1/\text{min}\lbrace\omega_{\alpha\beta}\rbrace$, $\omega_{\alpha\beta}=(E_\alpha-E_\beta)/\hbar$ the mean value approaches to $\langle\hat{\mathcal{A}}\rangle=\sum\limits_\alpha|c_\alpha|^2(\hat{\mathcal{A}})_{\alpha\alpha}$ meaning that contribution of the non-diagonal terms is marginal in the large time limit
$\sum\limits_{\alpha\neq\beta}c^*_\alpha c_\beta(\hat{\mathcal{A}})_{\alpha\beta}e^{-i\omega_{\alpha\beta}t} \rightarrow 0$. It is assumed that the dimension of the system's Hilbert space is finite but large. Currently, up-to-date cold atomic gases are considered a unique platform for testing the ETH hypothesis
\cite{deutsch2018eigenstate, PhysRevLett.112.130403,rigol2008thermalization,d2016quantum, PhysRevLett.110.257203, PhysRevLett.131.060401}.
In the present work, we propose an experimentally feasible system that could serve as an alternative platform when testing the fundamental principles of the ETH hypothesis. Namely, we consider a multimode optical system coupled to the JC system. We exploit the JC subsystem for quantum metrological purposes since, through the  JC subsystem and experimentally accessible level dynamics, we extract information about the thermalization of the multimode optical subsystem.

Thermodynamics in the last two decades has experienced progress, and the rise of new directions, such as non-equilibrium thermodynamics of small systems and quantum thermodynamics \cite{PhysRevResearch.2.043247, PhysRevE.101.020201, PhysRevLett.124.040602, PhysRevResearch.2.023377, PhysRevE.102.040103}. In the case of quantum systems, along with fluctuations in quantum thermodynamics, quantum effects play a dominant role \cite{PhysRevResearch.2.023120, kosloff2017quantum, PhysRevResearch.2.032062,delCampo2019, PhysRevLett.125.166802, PhysRevLett.124.110604, RevModPhys.91.045001, abah2017energy, PhysRevE.98.032121, azimi2014quantum, chotorlishvili2016superadiabatic, chotorlishvili2011thermal,PhysRevB.96.054440,PhysRevB.91.041408}.  Only recently have researchers been attracted to the thermodynamic theory of highly multimode nonlinear optical systems. In this context, we admit works published during the last few years    \cite{wu2019thermodynamic,marques2023observation,wright2022physics,wu2019thermodynamic}.
\begin{figure*}[ht]
    \centering
    \begin{minipage}{0.45\textwidth}
        \centering
        \includegraphics[width=\textwidth]{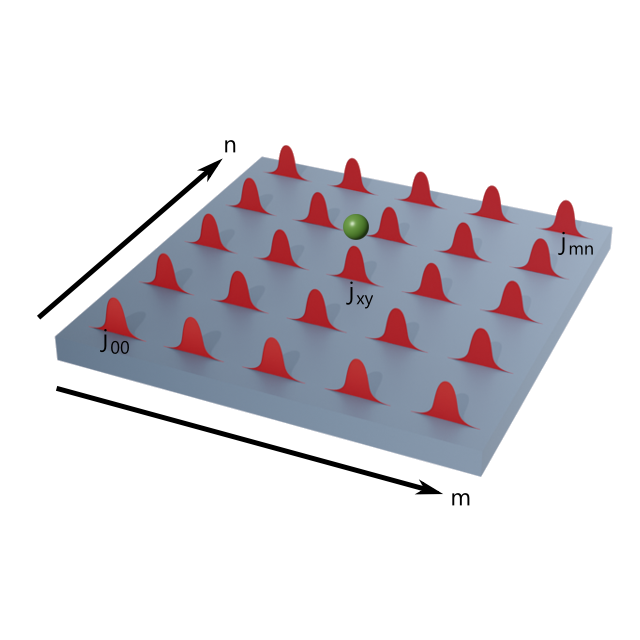}
    \end{minipage}
    \hfill
    \begin{minipage}{0.45\textwidth}
       \centering
        \includegraphics[width=\textwidth]{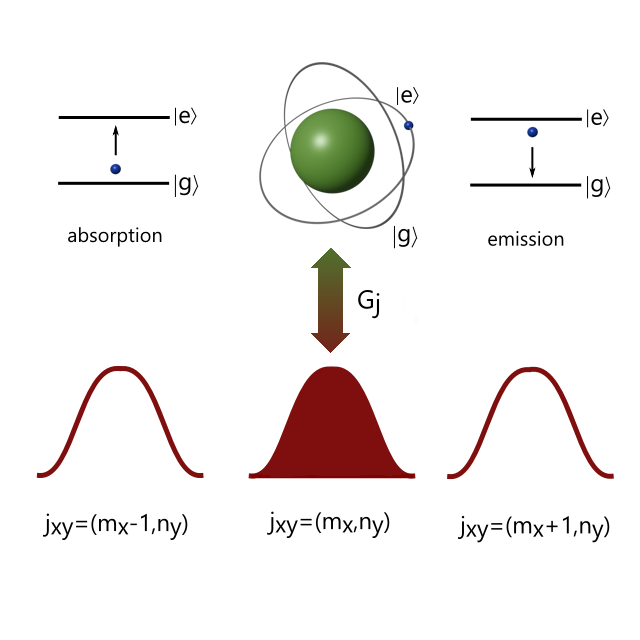}
    \end{minipage}
    \caption{Schematics of the proposed system. The multimode optical system is shown on the left side, where $j_{xy}$ corresponds to a single mode, with $n$ and $m$ being integers from $1$ to $N$, the total number of modes is equal to $N^2=400$. The interaction $G_j$ between Jaynes-Cummings two-level atoms with the particular mode is shown on the right side. $\ket{e}$ and $\ket{g}$ correspond to the excited and ground state of the atom. We are unaware to which particular mode atom is coupled. Therefore, we average the dynamics of level populations of the atom over the ensemble of all modes.}
    \label{fig:schematic}
\end{figure*}
Without claiming to be exhaustive, we briefly recall the peculiarities of thermodynamics of highly multimode nonlinear optical systems \cite{PhysRevLett.128.213901, PhysRevResearch.3.033219,wu2019thermodynamic,marques2023observation,wright2022physics,wu2019thermodynamic}.
The conserved internal energy of the optical multimode system is defined through the expectation value of the linear Hamiltonian $U=-H_L$
where $H_L=\langle\Psi\vert\hat H_L\vert\Psi\rangle$ and the state vector is the solution of the effective Schr\"odinger equation $id\ket{\psi}/dz=-\hat H_L\ket{\psi}$, $z$ being a spatial coordinate. The essence of the effective Schr\"odinger equation is that the propagation of wave along $\hat{\mathbf{z}}$ axis mimics the evolution in time. In what follows, we consider the conceptual problem of thermalization of a two-level atom coupled with the multimode optical system. For our goal, it is sufficient to consider a $N \times N$ square lattice (see Fig.\ref{fig:schematic}). The eigenmodes of the multimode optical system in the context of a square lattice of optical fibers with $\hat{\mathbf{z}}$ propagation direction reads:
\begin{equation}
\psi_j=c_je^{-i(\Omega t-k_0z-\varepsilon_jz)}
\label{modes}
\end{equation}
where $\varepsilon_j=\delta k(\cos p+\cos q)$ (with $\delta k\ll k_0$), $\Omega$ is a frequency of coherent photons, $k_0$ is an averaged propagation constant in square waveguide lattice, $\delta k$ is associated with overlap couplings between neighboring waveguides and $q$ and $p$ take the values $p=\pi n/N$ and $q=\pi m/N$. Here $n$ and $m$ are integers from $1$ to $N$ and thus the total number of modes $M=N^2$.
The expectation value of the Hamiltonian $\hat H_L$ farther can be presented in the form $H_L=\sum\limits_{j=1}^M\varepsilon_j|c_j|^2$. Here $|c_j|^2$ represents the mode occupation, and $\varepsilon_j$ is the corresponding eigenvalue.
The total power is given via $\mathcal{P}=\sum\limits_{j=1}^M|c_j|^2$. The nonlinear interactions between different modes lead to the thermalization of the system to the Rayleigh-Jeans distribution (for the detailed derivation, see Ref.\cite{wu2019thermodynamic}). The temperature $T$ in the model is introduced through the equation
\begin{eqnarray}\label{temperature is introduced}
|c_j|^2=-\frac{T}{\varepsilon_j+\mu},
\end{eqnarray}
where $\mu$ is the chemical potential. The equation of the state of the multimode system reads:
\begin{eqnarray}\label{The equation of the state}
&& U-\mu\mathcal{P}=MT,\nonumber\\
&& \mathcal{P}=-\sum\limits_{j=1}^M\frac{T}{\varepsilon_j+\mu},
\end{eqnarray}
where $M$ is the number of the mode.
We note that the internal energy of the multitude system and power are conserved quantities defined through the initial distribution of the amplitudes $U=-\sum\limits_{i=1}^M\varepsilon_i|c_{i0}|^2$, $\mathcal{P}=\sum\limits_{i=1}^M|c_{i0}|^2$.
Therefore, two equations Eq.(\ref{The equation of the state}) are enough to define two thermodynamic quantities, such as temperature $T$ and chemical potential $\mu$. In dimensionless units the total energy of the system is given by $E=\Omega\mathcal{P}$, where $\Omega$ is the frequency of photons. On the other hand $E=\alpha^2\hbar\Omega M$ where $\alpha^2$ is the mean photon number per mod estimated as $\alpha^2=\mathcal{P}/\hbar M$.  \vspace{0.2cm}\\
If the formation of the thermal state in the multimode optical system is beyond a doubt, then the two-level atom coupled to the multimode optical system should thermalize, too. In what follows, we exploit a two-level atom with experimentally observable level populations as a hyperfine sensor for the detection of thermalization. The two-level atom interacting with the particular $j$ mode is described by the Jaynes–Cummings model \cite{jaynes1963comparison}:
\begin{eqnarray}\label{JC Rabi 1}
&&\hat H_j=\hbar\Omega\hat a^+\hat a+\frac{1}{2}\hbar\omega\hat\sigma_z+G_j(z)(\hat\sigma\hat a^++\hat\sigma^+\hat a).
\end{eqnarray}
Here $\Omega$ is the frequency of coherent photons, $G_j(z)=\hbar g|c_j|\cos\left[(k_0+\varepsilon_j)z\right]$ describes the interaction of the two-level atom with the multimode system, $z$ is the distance of the two-level atom from the center of the $j$th mode. The particular mode $j$ here is chosen randomly. Therefore, we average the obtained solution over the ensemble of all $M=N^2=400$ modes and omit index $j$ in the equations for brevity. The states of the two-level atom are given by $\ket{e}=\begin{pmatrix}1\\0\end{pmatrix}$,  $\ket{g}=\begin{pmatrix}0\\1\end{pmatrix}$,  and $\hat\sigma^+=\ket{e}\bra{g}$, $\hat\sigma=\ket{g}\bra{e}$, and  $\hbar=1$ in what follows. To solve the Schr\"odinger equation
\begin{eqnarray}\label{Schrodinger}
i\frac{\partial\ket{\Psi}}{\partial t}=\hat H\ket{\Psi},
\end{eqnarray}
we assume that the initial state is the product state of the atom and field and the field is prepared in the coherent state \cite{Perelomov,Gilmore90} $\ket{\Psi}=\ket{\psi}_a\otimes\ket{\psi}_f=[c_e\ket{e}+c_g\ket{g}]\otimes\sum\limits_{n=0}^\infty w_n\ket{n}$, where $w_n=\frac{\alpha^m}{\sqrt{m!}}e^{-\alpha^2/2}$, $\ket{n}$ is the state with $n$ photons, and $|\alpha|^2$ is the mean photon number. From the solution of the Schr\"odinger equation
we construct the density matrix of the system $\rho_j(t)=\ket{\Psi}\bra{\Psi}$ and calculate the reduced density matrix of the atom \cite{PhysRevA.82.022110}:
\begin{eqnarray}\label{reducedone}
&&\hat\rho^j_a(t)=\text{Tr}_f(\hat\rho^j(t))=\sum\limits_n\bra{n}\hat\rho^j(t)\ket{n},
\end{eqnarray}
$\text{Tr}_f$ means the trace over the states of photonic field.
In the explicit form:
\begin{eqnarray}\label{1Bassano Vacchini}
&&\rho^j_{11}(t)=\rho_{00}(0)[1-\alpha(t)]+\rho_{11}(0)\beta(t),\nonumber\\
&&\rho^j_{00}(t)=\rho_{00}(0)\alpha(t)+\rho_{11}(0)[1-\beta(t)],\nonumber\\
&&\rho^j_{10}(t)=\rho_{10}(0)\gamma(t)\nonumber\\
&&\rho^j_{01}(t)=\rho_{01}(0)\gamma^*(t).
\end{eqnarray}
Here we introduced the notations:
\begin{eqnarray}\label{2Bassano Vacchini}
&&C_n(t)=e^{i(\omega-\Omega)t/2}\bigg[\cos(\lambda_nt/2)-\nonumber\\
&&i\frac{(\omega-\Omega)}{\lambda_n}\sin(\lambda_nt/2)\bigg],\nonumber\\
&&\alpha(t)=\sum_{n}C^*_{n}(t)C_{n}(t),\nonumber\\
&&\beta(t)=\sum_{n}C^*_{n+1}(t)C_{n+1}(t),\nonumber\\
&&\gamma(t)=\sum_nC_{n}(t)C_{n+1}(t),
\end{eqnarray}
and the expression of the Rabi frequency has the form  $\lambda_n=\sqrt{(\omega-\Omega)^2+4(n+1)G^2_j(z)}$, $G_j(z)=\hbar g|c_j|\cos\left[(k_0+\varepsilon_j)z\right]$.
We are interested in the diagonal matrix elements, which we define in terms of atomic inversion after averaging over the ensemble of modes:
\begin{eqnarray}\label{inversion one}
&&\rho_{11}=\frac{1}{2}\left(1+\mathcal{W}\right),\nonumber\\
&&\rho_{22}=\frac{1}{2}\left(1-\mathcal{W}\right).
\end{eqnarray}
Following \cite{karatsuba2009resummation} we present atomic inversion in the form:
\begin{eqnarray}\label{inversion two}
&&\mathcal{W}=\mathcal{A}_0+\mathcal{A}(t),\nonumber\\
&&\mathcal{A}_0=\sum\limits_{n=0}^\infty\sum\limits_{j=1}^M\frac{|\alpha|^{2n}e^{-|\alpha|^2}}{M\cdot n!}\frac{(\omega-\Omega)^2}{\lambda_n^2},\nonumber\\
&&\mathcal{A}(t)=\sum\limits_{n=0}^\infty\sum\limits_{j=1}^M\frac{|\alpha|^{2n}e^{-|\alpha|^2}}{M\cdot n!}\frac{2(n+1)(\hbar g)^2}{\lambda_n^2}\times\nonumber\\
&&|c_j|^2\left(1+\cos[2(k_0+\epsilon_j)z]\right)\cos(\lambda_nt).
\end{eqnarray}
It is easy to see that only the term $\mathcal{A}(t)$ depends on the time. Therefore, the study of the thermalization process is reduced to the asymptotic limit of $\mathcal{A}_{t\rightarrow\infty}$.
After performing summation for over the $n$ we deduce:
\begin{eqnarray}\label{inversion three}
\mathcal{A}_0=\sum\limits_{j=1}^M\frac{(\omega-\Omega)^2}{(\omega-\Omega)^2+4|\alpha|^2G_j^2(z)},
\end{eqnarray}
while for the time-dependent part, we deduce
\begin{eqnarray}\label{inversion four}
&&\mathcal{A}(t)=\sum\limits_{n=0}^\infty\frac{2(\hbar g)^2(n+1)|\alpha|^{2n}e^{-|\alpha|^2}}{M\cdot n!}\mathcal{F}_n(t),\nonumber\\
&&\mathcal{F}_n(t)=\sum\limits_{j=1}^M\frac{|c_j|^2}{\lambda_n^2}\left(1+\cos[2(k_0+\varepsilon_j)z]\right)\cos(\lambda_nt).
\end{eqnarray}
The effective atomic temperature is given by
\begin{eqnarray}\label{atomic temperature}
T_a=\frac{\hbar\omega}{k_B\ln\left[\frac{1+\mathcal{W}}{1-\mathcal{W}}\right]}.
\end{eqnarray}
In the high temperature limit all modes are equally populated, meaning that Eq.(\ref{temperature is introduced}) simplifies as follows: $|\mu|\gg\varepsilon_j$,
$|c_j|^2=T/|\mu|=$const, with $T\gg1$. Then summation over the ensemble of modes in Eq.(\ref{inversion three}) and Eq.(\ref{inversion four}) can be performed analytically and we deduce:
\begin{eqnarray}\label{performed analytically}
&&\mathcal{W}(t)=\frac{e^{-\alpha^2}\left(1+\cos(\lambda_0t)\right)\varepsilon_{max}}{2M}-\nonumber\\
&&\frac{e^{-\alpha^2}\left(1-\cos(\lambda_0t)\right)\varepsilon_{max}}{2Mz\sqrt{2\xi+1}}\bigg(\arctan\bigg[\frac{\tan[z\varepsilon_{max}+k_0]}{\sqrt{2\xi+1}}\bigg]-\nonumber\\
&&\arctan\bigg[\frac{\tan[k_0]}{\sqrt{2\xi+1}}\bigg]\bigg),
\end{eqnarray}
where we introduced the notation $\xi=\frac{2(\hbar g)^2|c|^2}{(\omega-\Omega)^2}$. From Eq.(\ref{performed analytically}) it is easy to see that $\mathcal{W}(t)\rightarrow 0$ as $M\gg1$ and $\alpha^2\gg1$. Then temperature of the atom approaches to the temperature of the modes:
\begin{eqnarray}\label{approaches to}
&&T_a=\frac{\hbar\omega}{k_B\mathcal{W}}\rightarrow\infty,\,\,\,T\rightarrow\infty.
\end{eqnarray}
\begin{table}\label{Table}
	\caption{The value of $A_0$ (in bold phase) for different values of power $P$, internal energy $U$, chemical potential $\mu$, the value of the mode temperature $T$ and the mean photon number $|\alpha|^2$. The number of the modes $M = N^2 = 400$.\\}
    \centering
    \begin{tabular}{p{0.8cm} | p{1.3cm} | p{1.3cm} | p{1cm} | p{1cm} | p{1cm} | p{1cm} }
		\multirow{2}{*}{$P$} & \multirow{2}{*}{$U$} & \multirow{2}{*}{$\mu$} & \multirow{2}{*}{$T$} & \multicolumn{3}{c}{$\left|\alpha\right|^2$}\\\cline{5-7}
		& & & & 10 & 20 & 50 \\\hline
		$200$& $75.689$&$-7.266$&$3.822$ & \bf{0.182} & \bf{0.127} & \bf{0.077} \\\hline
		$300$& $109.750$&$-7.020$&$5.539$ & \bf{0.147} & \bf{0.102} & \bf{0.061} \\\hline
		$400$& $141.250$&$-6.766$&$7.117$ & \bf{0.126} & \bf{0.087} & \bf{0.052} \\
	\end{tabular}
\end{table}
\begin{figure}
	\includegraphics[width=0.95\columnwidth]{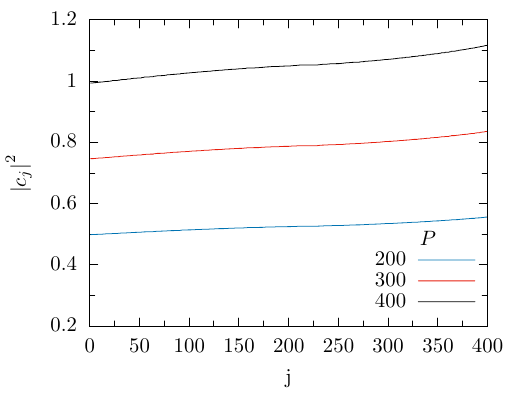}
	\caption{The stored distribution of the mode amplitudes $\left|c_j\right|^2$. Slight increase of the amplitudes with $j$ is a signature of negative chemical potential of the modes. }
    \label{fig:cj2}
\end{figure}
\begin{figure}
	\includegraphics[width=0.95\columnwidth]{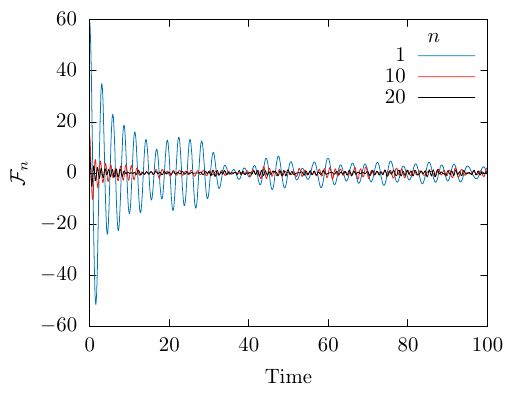}
	\caption{Decay of partial contributions$\mathcal{F}_n(t)$ in Eq.(\ref{inversion four}) related to the states with a particular number of photons $\ket{n}$, for different $n$ and mean photon number $\left|\alpha\right|^2 = 10$. Initial power is equal to $\mathcal{P} = 200$.}
     \label{fig:Fn}
\end{figure}
\begin{figure}
	\includegraphics[width=0.95\columnwidth]{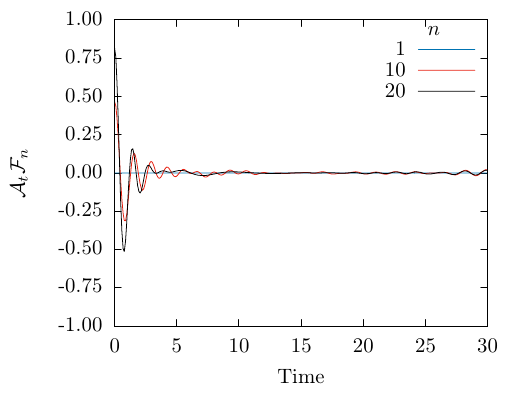}
	\caption{The time-dependent part of the level inversion Eq.(\ref{inversion four}) $\mathcal{A}(t)$ summed up for the states from the single $1$ up to $n=20$ photons. The mean photon number is equal to $\left|\alpha\right|^2 = 10$, $P = 200$.}
    \label{fig:AtFn}
\end{figure}

\textbf{Results and conclusions:} For the finite mode temperature case, the atom temperature is distinct from the mode temperature. In Fig.\ref{fig:cj2}, we plot the distribution of the mode amplitudes $|c_j|^2$ after thermalization. Amplitudes are sorted in the order of the modes
$\varepsilon_{j+1}>\varepsilon_j$.  As we see, the values of $|c_j|^2$ are increasing with $j$ meaning the negative chemical potential of the optical multimode system. With increasing of power $\mathcal{P}$, i.e., the energy pumped in the optical modes, the distribution of the amplitudes becomes more uniform because  of $T\gg\text{max}(\varepsilon_j)$, $|\mu|\gg\text{max}(\varepsilon_j)$. In Table \ref{Table}, we present results for the time-independent part of the inversion $\mathcal{A}_0$,  Eq.(\ref{inversion two}) calculated for the different values of parameters. As we see from the table, with the increasing mode temperature $T$ and mean photon number $|\alpha|^2$, value of  $\mathcal{A}_0$ tends to zero, which is a signature of the infinite temperature of the atom (both levels are equally populated). The thermalization process of the atom is described by the time-dependent part of inversion Eq.(\ref{inversion four}) and is plotted in Figs. \ref{fig:Fn}, \ref{fig:AtFn}. As we see from the plots, the dependent part of the inversion decays in time, and instead of the quantum revivals, we see small fluctuations only. Function $\mathcal{F}_n(t)$ in Fig. \ref{fig:Fn} characterizes decay of partial contributions in Eq.(\ref{inversion four}) related to the states with a particular number of photons $\ket{n}$, while $\mathcal{A}_n(t)$ in Eq.(\ref{fig:AtFn}) describes the effect of all states.
In the present project, we study the dynamics of the two-level system coupled with the multimode optical system. In particular, we considered a square lattice of optical fibers. We aimed to answer whether we can infer information about the thermalization of optical modes through the thermalization of two-level atoms. After averaging over the set of modes, the dynamic of the two-level system is free of quantum revivals, and that is the signature of thermalization. We showed that the temperature of the two-level system increases with the temperature of optical modes and mean photon number. In the high-temperature limit of optical modes, the temperature of the level system tends to be infinity, and level populations are equal.

\bibliography{reference}

\end{document}